\documentclass[twocolumn,aps,showpacs,prl,amsmath,amssymb,floatfix,superscriptaddress]{revtex4-2}
 
\usepackage{amssymb,color}

\usepackage{bm} 
\usepackage{color}
\usepackage{graphicx}
\usepackage{physics}
\usepackage{amsthm}
\usepackage{amsmath}
\usepackage{amssymb}
\usepackage{enumerate}
\usepackage{placeins}
\usepackage{booktabs}
\usepackage{dsfont}
\usepackage{yfonts}
\usepackage{hyperref} 
\usepackage{float}

\DeclareSymbolFont{bbold}{U}{bbold}{m}{n}
\DeclareSymbolFontAlphabet{\mathbbold}{bbold}

\newcommand{\e}{{\rm e}}  
\newcommand{\ex}[1]{\langle #1 \rangle}

\usepackage{pdfpages} 
\usepackage{pgffor} 

\makeatletter
\AtBeginDocument{\let\LS@rot\@undefined}
\makeatother

\def\supplementfilename{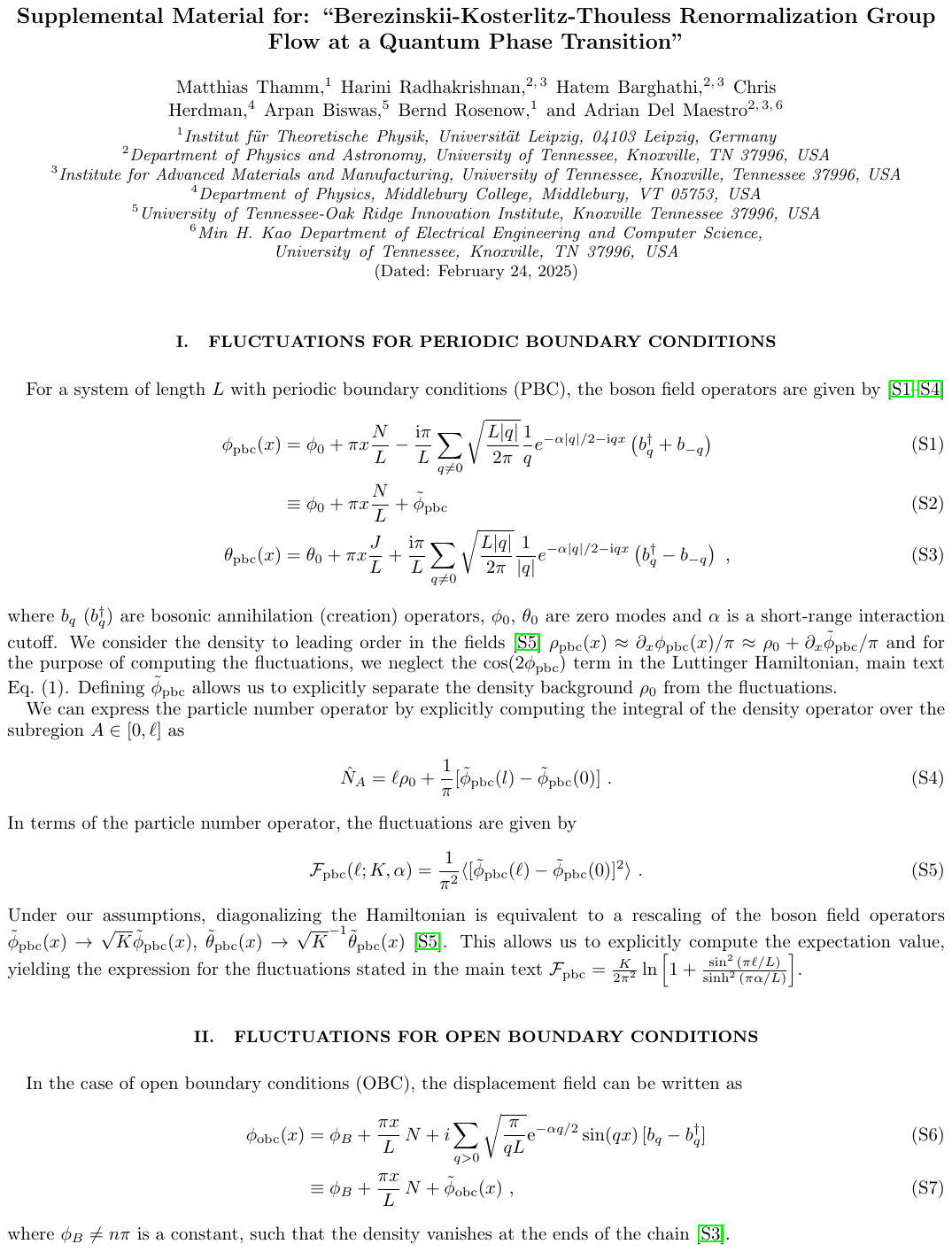}

\pdfximage{\supplementfilename}
\def\numbersupplementpages{\the\pdflastximagepages}

\newif\ifarXiv
\arXivtrue 

\begin{document}
\title{Berezinskii-Kosterlitz-Thouless Renormalization Group Flow at a Quantum Phase Transition} 

\author{Matthias Thamm}
\affiliation{Institut f\"{u}r Theoretische Physik, Universit\"{a}t Leipzig, 04103 Leipzig, Germany}

\author{Harini Radhakrishnan}
\affiliation{Department of Physics and Astronomy, University of Tennessee, Knoxville, TN 37996, USA}
\affiliation{Institute for Advanced Materials and Manufacturing, University of Tennessee, Knoxville, Tennessee 37996, USA\looseness=-1}

\author{Hatem Barghathi}
\affiliation{Department of Physics and Astronomy, University of Tennessee, Knoxville, TN 37996, USA}
\affiliation{Institute for Advanced Materials and Manufacturing, University of Tennessee, Knoxville, Tennessee 37996, USA\looseness=-1}

\author{C. M. Herdman}
\affiliation{Department of Physics, Middlebury College, Middlebury, VT 05753, USA}

\author{Arpan Biswas}
\affiliation{University of Tennessee-Oak Ridge Innovation Institute, Knoxville Tennessee 37996, USA}

\author{Bernd Rosenow}
\affiliation{Institut f\"{u}r Theoretische Physik, Universit\"{a}t Leipzig, 04103 Leipzig, Germany}

\author{Adrian Del Maestro}
\affiliation{Department of Physics and Astronomy, University of Tennessee, Knoxville, TN 37996, USA}
\affiliation{Institute for Advanced Materials and Manufacturing, University of Tennessee, Knoxville, Tennessee 37996, USA\looseness=-1}
\affiliation{Min H. Kao Department of Electrical Engineering and Computer Science, University of Tennessee, Knoxville, TN 37996, USA}

\date{\today}

\begin{abstract}
We present a controlled numerical study of the Berezinskii-Kosterlitz-Thouless (BKT) transition in the one-dimensional Bose-Hubbard model at unit filling, providing evidence of the characteristic logarithmic finite-size scaling of the BKT transition.  
Employing density matrix renormalization group and quantum Monte Carlo simulations under periodic boundary conditions, together with a systematic finite-size scaling analysis of bipartite particle number fluctuations, we resolve boundary-induced complications that previously obscured critical scaling.
 We demonstrate that a suitably chosen central region under open boundaries reproduces universal RG signatures, reconciling earlier discrepancies. Finally, leveraging a non-parametric Bayesian analysis, we determine the critical interaction strength with high precision, establishing a benchmark for BKT physics in one-dimensional quantum models.
 \end{abstract}

\maketitle
\emph{Introduction---}The Berezinskii-Kosterlitz-Thouless (BKT) transition \cite{Berezinskii.1971,Kosterlitz.1973} is a remarkable infinite-order topological phase transition, originally discovered in the classical two-dimensional (2D) XY model. 
The low-temperature phase is characterized by algebraically decaying spin correlations, 
and the transition to a high-temperature disordered phase is driven by the unbinding of topological defects (vortex-antivortex pairs)  due to thermal fluctuations. The location of the critical point is known from Monte Carlo \cite{Weber1988,Hsieh2013,Sandvik2006,Zhang2019,Nguyen2021} and tensor network simulations \cite{Vanderstraeten2019,Jha2020,Ueda2021}. 
BKT physics has been found in a variety of systems, including the 2D Coulomb gas \cite{minnhagen1987two}, superconducting films \cite{fiory1983superconducting,matthey2007electric,kamlapure2010measurement,rout2010interface,goldman2013berezinskii,mironov2018charge}, Josephson junction arrays \cite{capriotti2004berezinskii,bottcher2022berezinskii}, superconductor-ferromagnetic-superconductor junctions \cite{reich:2024bk}, and lattice models at zero temperature \cite{hadzibabic2006berezinskii,castelnovo2007zero}.
Near the transition, all these systems are expected to follow a renormalization group (RG) flow characteristic of the BKT universality class.
Although this RG flow has been numerically demonstrated in the classical XY model \cite{Ueda2021,Melko2004,MasakiKato2019,Schaffer2009,Owerre2013,shao2020monte}, 
its observation in quantum systems (with the transition driven  by quantum instead of thermal fluctuations) remains a major challenge.  This is due to a signature feature of BKT scaling -- logarithmic system size dependence at criticality -- necessitating costly large-scale numerical simulations of quantum systems.

\begin{figure*}[t!]
    \centering
    \includegraphics[width= \linewidth]{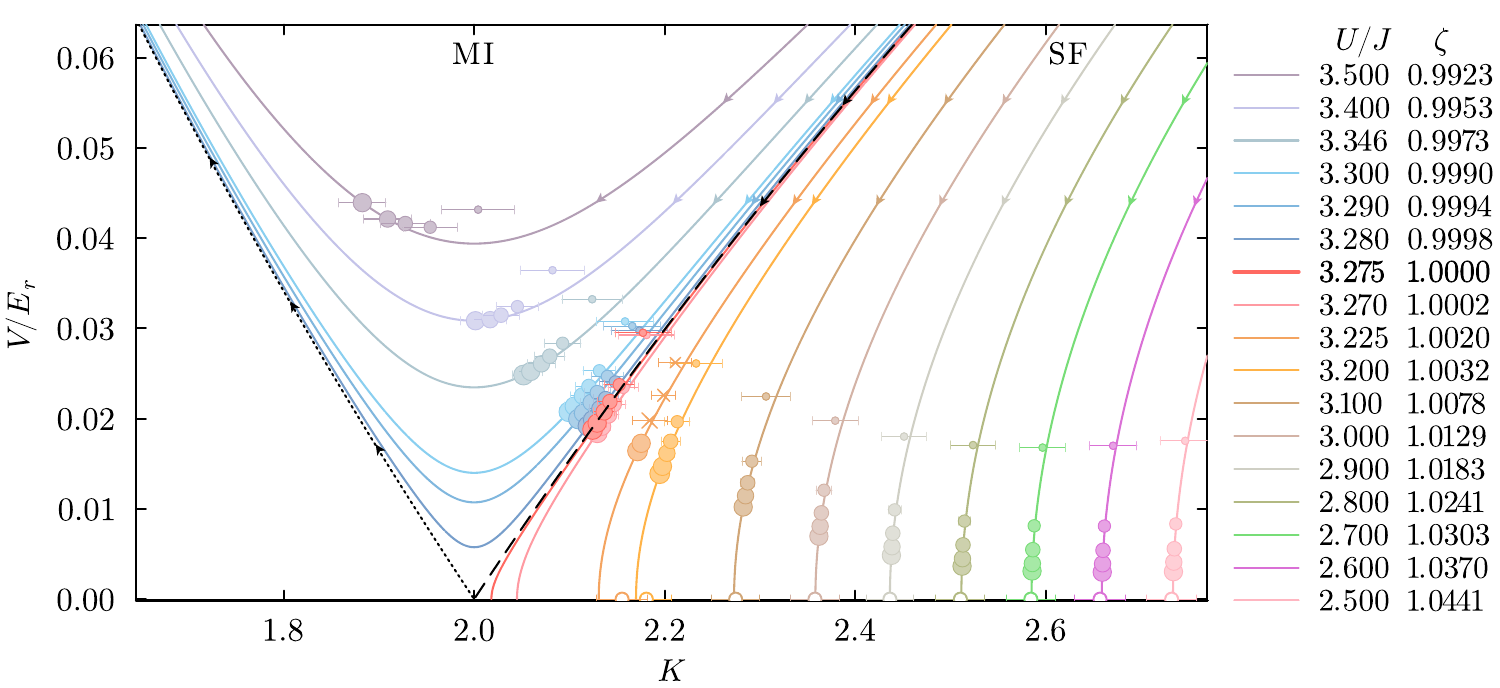}
    \caption{RG flow close to the BKT phase transition constructed from the numerical DMRG simulations for periodic boundary conditions via Eq.~\eqref{Eq:RGEquation}. Here, $K$ is the Luttinger parameter, $V$ the strength of the periodic lattice potential, $E_r=u\pi/a$ a characteristic energy scale, and $U$ the interacting strength of the Bose-Hubbard model, Eq.~\eqref{Eq:HBH}, considered in the numerical simulations. The dashed black line is the separatrix marking the critical line of the phase transition with $\zeta(U_c/J)=1$. To the right of the separatrix, the system is in the superfluid phase (SF),  and to the left  it is in the Mott insulating phase (MI). Each colored line corresponds to one value of $\zeta(U/J)$ obtained from the numerical results for the largest two system sizes for each $U/J$.  Circles show the points for individual system sizes (larger circles correspond to larger $L$, $L\in\{16,32,48,64,96,128\}$).  Open circles are obtained by fits of the $L\to\infty$ limit of Eq.~\eqref{Eq:Fpbc} to data obtained for infinite boundary conditions using the VUMPS algorithm \cite{ZaunerStauber.2018}. Crosses are obtained from QMC. We estimate error bars on $K$ by varying the fit intervals.  We find excellent agreement of the DMRG data with the predicted RG flow and between the different methods.}
    \label{Fig:RGflowPBCdata}
\end{figure*}

The transition between superfluid and Mott insulator in the one-dimensional (1D) Bose-Hubbard (BH) model is a paradigmatic example of  quantum BKT physics, corresponding to a 2D classical BKT transition via the  quantum-to-classical mapping.
The transition is governed by a competition between the kinetic energy of the bosons and the on-site repulsive interaction $U$.
For small interactions, the system is in a  superfluid phase, while it is in a Mott insulating phase when the on-site repulsion is dominant. 
This quantum phase transition has been experimentally observed in cold-atom systems \cite{Esslinger:2004tf,haller2010}
and due to its intrinsic interest, a variety of theoretical studies have attempted to locate the critical point in the 1D BH model \cite{krutitsky2016ultracold}.  This includes approximate analytical approaches \cite{Krauth.1991}, exact diagonalization \cite{Elesin.1994}, density matrix renormalization group (DMRG)  \cite{Kuhner:1998bh,Roux.2008,Rossini.2012,Laeuchli.2008,Ejima.2011,Rachel.2012,carrasquilla2013scaling}, infinite matrix product state simulations \cite{cincio2019universal,rams2018precise}, quantum Monte Carlo (QMC) \cite{Batrouni.1992,Gerster2016,Kashurnikov.1996,Bergkvist2007}, time evolving block decimation (TEBD) \cite{Zakrzewski.2008,Danshita.2011,Pino.2012,Zuo2021}, and entanglement-based techniques \cite{Contessi.2023}. While these studies have reported a range of critical interaction strengths with varying degrees of accuracy and precision, the expected BKT RG flow behavior remains unobserved.

A key difficulty in pinpointing the BKT RG flow in quantum simulations arises from boundary conditions. For open boundary conditions (OBC), a position-dependent vortex fugacity and superfluid stiffness must be introduced to account for ``image vortices" induced by the system’s edges \cite{Artemov:20052d,Holz:1987bc,Falo:1989mc}, complicating direct comparisons to RG predictions developed for infinite or periodic systems.  This is a non-trivial boundary effect, with the emergent low-energy description taking on the form of a non-local Sine-Gordon model \cite{Artemov:20052d}. Capturing these subtleties is essential for extracting BKT scaling, and their neglect may explain the spread in critical parameters observed in DMRG studies with OBC \cite{Kuhner:1998bh,Roux.2008,Rossini.2012,Laeuchli.2008,Ejima.2011,Rachel.2012}.

In this Letter, we perform a  theoretical analysis of the BKT transition in the 1D Bose-Hubbard model by combining periodic boundary conditions (PBC) DMRG  with a systematic finite-size scaling analysis of bipartite particle number fluctuations \cite{Rachel.2012}. Our main contributions are: (i) an observation of the BKT RG flow at a quantum phase transition, overcoming the boundary-related issues that affected previous studies; (ii) when using OBC, identifying the small fraction of the system near the center that can be used to extract observables compatible with infinite size RG predictions; 
and (iii) a high-precision determination of the critical interaction strength in the Bose-Hubbard model at unit filling with uncertainty quantification via a non-parametric Gaussian process model.

\emph{Model---}We consider a one-dimensional system of interacting quantum particles at zero-temperature described by the Luttinger liquid Hamiltonian \cite{Giamarchi.2010}
    \begin{align}
         H_{\rm LL} &= \int_0^L\!\!\! {\rm d}x\,  \frac{u}{2\pi}\left[ K(\partial_x\theta)^2 + \frac{1}{K} (\partial_x\phi)^2\right] + \frac{V}{2\pi a}\cos(2\phi) \ , \label{Eq:LLHam}
    \end{align}
with bosonic phase field $\theta(x)$, displacement field $\phi(x)$, and a pinning term due to a periodic lattice potential of strength $V$. Here, $u$ is the velocity of excitations, $a$ the lattice spacing, and $K>1$ is the Luttinger parameter determined by the strength of repulsive interparticle interactions. 
In the low energy limit under the assumption of an average density $\rho_0$ and excitations causing small fluctuations around $\rho_0$, the particle field operator $\psi(x)$ is related to the phase field via $\psi(x)\approx \sqrt{\rho(x)}\e^{i\theta(x)}$, and the density profile is given by 
$\rho(x)\approx \partial_x\phi(x)/\pi \equiv \rho_0+\partial_x\tilde{\phi}(x)/\pi $  \cite{Giamarchi.2010,Mattsson.1997,Eggert.1992,Cazalilla.2004}.

The Hamiltonian \eqref{Eq:LLHam} has the form of a sine-Gordon model \cite{Giamarchi.2010}, which is known to host a Berezinskii-Kosterlitz-Thouless (BKT) phase transition \cite{Berezinskii.1971,Kosterlitz.1973} at $K(L\to\infty)=2$ close to which the RG flow is governed by the flow equations 
$\frac{{\rm d}K}{{\rm d}\ell } = -\left(\frac{\pi V}{E_r}\right)^2\frac{K^2}{4}$ and  $\frac{{\rm d}V}{{\rm d}\ell } = (2-K) V$ \cite{Giamarchi.2010} with a characteristic energy scale $E_r=u\pi/a$ and $\ell = \ln L$ sets the RG scale.

The flow equations can be solved by the introduction of $\zeta(K,V)$ with ${\rm d}\zeta/{\rm d}\ell =0 $ \cite{Boeris.2016}\footnote{The flow equations stated in the supplemental material of Ref.~\cite{Boeris.2016} contain a typo that is corrected in Eqs.~\eqref{Eq:RGEquation} and \eqref{Eq:RGZeta}.}, resulting in 
\begin{align}
    2\ln\!\left(\frac{L_2}{L_1}\right)  &= \int_{K_2}^{K_1} \frac{{\rm d}K}{K^2 (\ln(K/2)-\zeta) + 2K} \label{Eq:RGEquation}\\
            \zeta  &\equiv \frac{2}{K} + \ln\left(\frac{K}{2}\right) - \frac{\pi^2}{2} \left(\frac{V}{E_r}\right)^2 \label{Eq:RGZeta} \ .
\end{align}
Given the values of $(K_1,K_2)$ for two system sizes $(L_1,L_2)$, solving Eq.~\eqref{Eq:RGEquation} for $\zeta$ determines a line in the flow diagram $K(V;L)$ describing how $K$ and $V$ change when increasing the system size $L$ (solid lines in Fig.~\ref{Fig:RGflowPBCdata}). Knowledge of the flow line allows for the finite-size scaling of $K$ to the thermodynamic limit $L\to\infty$. In the superfluid phase (SF), $\zeta>1$, the Luttinger parameter $K$ flows to a line of fixed points $K^*(\zeta)>2$ and $V$ to zero. 
The separatrix (dashed black), i.e.\@ the critical flow line of the quantum phase transition, corresponds to $\zeta=1$. For strong interactions, $\zeta<1$, the system is in the Mott insulating (MI) phase where $K$ flows to zero and $V$ to infinity.

\begin{figure}[t!]
    \centering
    \includegraphics[width=1.03\linewidth]{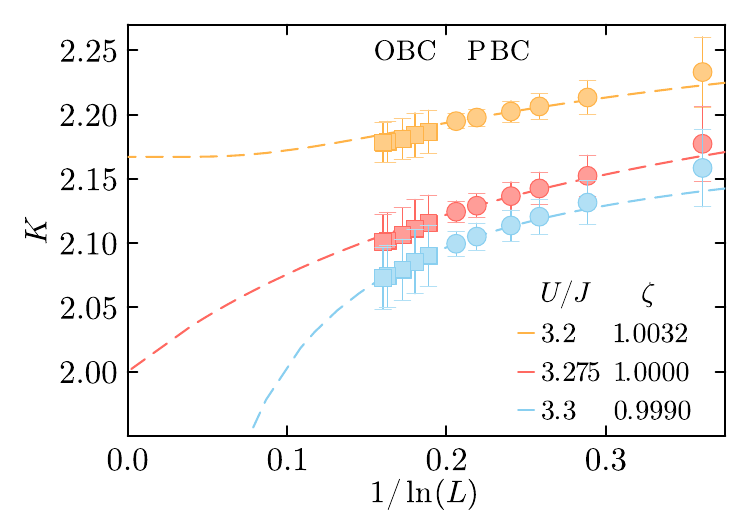}
    \caption{Finite size scaling of the Luttinger parameter as a function of the system size $L$ according to the RG flow. The circles depict the results obtained from high precision DMRG calculations with open boundary conditions. Dashed lines show fits of the corresponding scaling form to the data. We show the three scaling regimes (i) in the superfluid phase according to Eq.~\eqref{Eq:ScaleSF} (upper orange curve), (ii) for the value of $U$ closest to the separatrix using Eq.~\eqref{Eq:ScaleSEP} (center red curve), and (iii) in the insulating phase according to Eq.~\eqref{Eq:ScaleINS} (bottom blue curve). Squares depict data obtained for open boundary conditions using a finetuned fitting interval for extraction of $K$ \cite{Supplement}.}
    \label{Fig:K_of_L_FSS}
\end{figure}

\emph{RG flow finite size scaling---}Based on the RG flow, we can approximate the finite size scaling form of the Luttinger parameter $K$ analytically in three regions of the BKT phase diagram:  
\begin{align}
        K_{\rm SF} &\sim K^* + \kappa_0 \left(\frac{L_0}{L}\right)^{2(K^*-2)} \label{Eq:ScaleSF}   \\
        K_{\rm sep} &\sim 2  + \frac{1}{1/\kappa_0 + \ln L/L_0} 
    \label{Eq:ScaleSEP}   \\ 
      K_{\rm MI} &\sim 1 + (1+\kappa_0)\left(\frac{L_0}{L}\right)^{8 (1-\zeta)}\, . \label{Eq:ScaleINS}  
\end{align}
For weak lattice potential strength $V$ in the superfluid (SF) phase, we expand 
$K_{\rm SF} = K^*+\kappa$ for small $\kappa$ resulting in the scaling Eq.~\eqref{Eq:ScaleSF}. Here, $K(L_0)=K^*+\kappa_0$ for some system finite size $L_0$ with distance $\kappa_0$ from $K^*$.  For the critical flow on the separatrix (sep), $\zeta=1$ and $K^*=2$, yielding the finite size scaling Eq.~\eqref{Eq:ScaleSEP}. In the insulating phase, as $K$ flows to zero, one cannot expand for the asymptotic $L$ dependence. However, $K(V)$ has a minimum at $K=2$ close to which the scaling form is given by Eq.~\eqref{Eq:ScaleINS}.

This shows that the finite size scaling is logarithmic on the separatrix and nearly-logarithmic, i.e.\@ of power-law type with small exponent, close to it. 
Thus the functional form of the finite size scaling (even asymptotically) depends on the exact point in the phase diagram. This renders typical empirical $1/L$ extrapolations for $L\to\infty$ inaccurate, and requires utilization of the RG flow to produce accurate predictions of the thermodynamic limit behavior from finite size numerical simulations.  Examples of the scaling forms in Eqs.~\eqref{Eq:ScaleSF}--\eqref{Eq:ScaleINS} are shown in Fig.~\ref{Fig:K_of_L_FSS} (dashed lines), together with numerical data obtained for the Bose-Hubbard model (circles, PBC; squares OBC) as described below.

\emph{Bose-Hubbard model---}As an example system, we study the Bose-Hubbard model at zero temperature
\begin{align}
    H &= -J \sum_{i=1}^L \left( b_i^\dagger b^{\phantom{\dagger}}_{i+1}+ b_{i+1}^\dagger b^{\phantom{\dagger}}_i \right) + \frac{U}{2} \sum_{i=1}^L n_i (n_i-1) \label{Eq:HBH}
\end{align}
on a one dimensional, periodic lattice of length $L$. Here, $J$ is the nearest neighbor hopping strength, $U>0$ the repulsive onsite interaction, $b_i\, (b_i^\dagger)$ annihilates (creates) a boson on site $i$, and $n_i=b_i^\dagger b_i$ is the number operator for site $i$. For PBC, we identify $b_{L+1}=b_1$, while for OBC, $b_{L+1}=0$. We consider unit filling where $\expval{n_i} = N/L = 1$.

The low energy sector of Eq.~\eqref{Eq:HBH} can be described by the Luttinger liquid Hamiltonian \eqref{Eq:LLHam} via bosonization, which allows for the relation of the interaction strength $U$ to the Luttinger parameter $K$. Here, the density pinning term with pre-factor $V/E_r$ is related to Umklapp processes on the lattice \cite{Giamarchi:1997mt} and becomes interaction dependent.  The Bose-Hubbard model thus exhibits a BKT phase transition between a Mott-insulating phase for $U>U_c$ ($K<2$) and superfluid phase for $U<U_c$ ($K>2$) and falls under the universality class of models following the RG flow described above \cite{Haldane:1981wm} and depicted in Fig.~\ref{Fig:RGflowPBCdata}. The critical point can then, in principle, be determined by inverting the unknown function $K(U/J)$.  

\begin{figure}[t!]
    \centering
    \includegraphics[width=\linewidth]{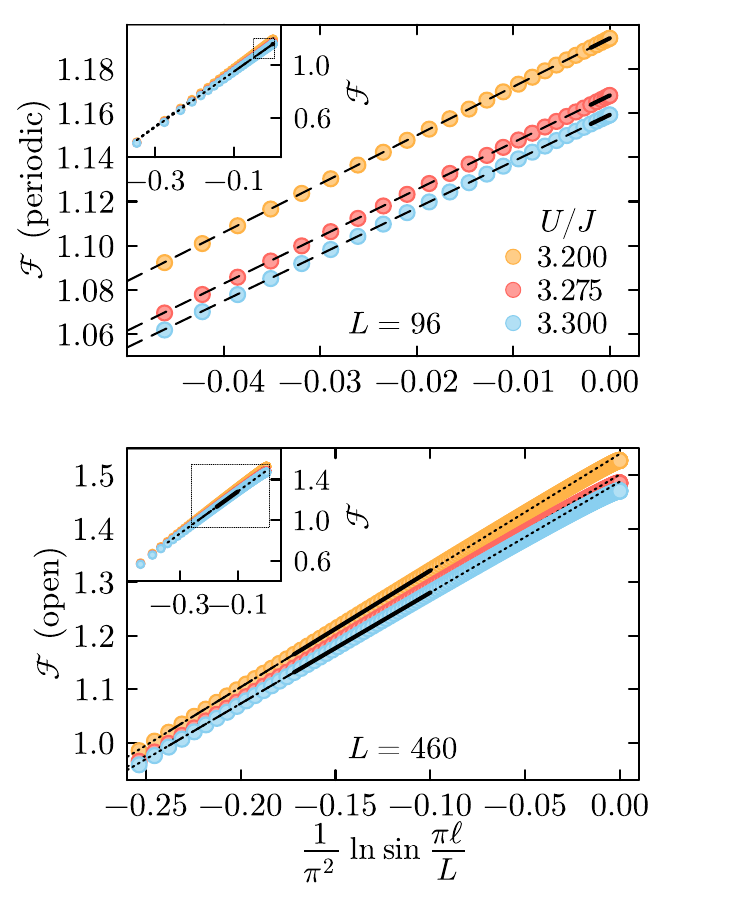}
    \caption{Extracting the Luttinger parameter $K$ from the particle number fluctuations for different interaction strengths $U/J$. We fit Eq.~\eqref{Eq:Fpbc} to the periodic boundary condition DMRG data (upper panel) for intervals close to $\ell=L/2$, i.e.\@ $\ell\in[L/2-L/16,L/2]$, with the extracted value of $K$ being insensitive to the choice of fitting region. The bottom panel shows fits to the open boundary condition data, where we fit to a small center region $\ell\in[L/8.2-L/16,L/8.2]$ such that $\ell$ covers the same absolute range as in the top panel. This restriction is necessary to mitigate the strong  boundary effects for $\ell$ approaching $L/2$ which leads to a strong dependence of $K$ on the fitting region, ultimately causing a large uncertainty in the extracted value.}
    \label{Fig:F_of_K_fit}
\end{figure}

We extract $K$ from the fluctuations of the particle number in an interval $A=[0,\ell]$ defined as  $\mathcal{F}(\ell)  = \ex{(\hat{N}_A - \ex{\hat{N_A}})^2}$ \cite{Rachel.2012}, where the particle number operator is related to the density via $ \hat{N}_A = \int_0^\ell dx\;\rho(x)$. Particle number fluctuations are proportional to the Luttinger parameter, and for PBC are given by
\begin{align}
    \mathcal{F}_{\rm{pbc}}(\ell; K, \alpha) &= \frac{1}{\pi^2} \langle[\tilde{\phi}(\ell)-\tilde{\phi}(0)]^2\rangle \\ 
    &= \frac{K}{2\pi^2}\ln{\left[1+\frac{\sin^2{(\pi\ell/L)}}{\sinh^2{(\pi\alpha/L)}}\right]} 
    \label{Eq:Fpbc}
\end{align}
with $\alpha$ a short-distance cutoff. The OBC case is discussed in the supplement \cite{Supplement}. For the region $\ell\sim L/2$ in the case of sufficiently large systems, such that $ \ell/L\gg \alpha/L$, we can write $\mathcal{F}_{\rm{pbc}}(\ell; K, \alpha) \approx \frac{K}{\pi^2} \ln \sin\frac{\pi\ell}{L} + A$ with constant $A\in\mathbb{R}$, allowing us to extract $K$ as the slope of a linear fit to $\mathcal{F}_{\rm pbc}$ plotted against $\frac{1}{\pi^2}\ln \sin\frac{\pi\ell}{L}$ as seen in Fig.~\ref{Fig:F_of_K_fit}.

\emph{Microscopic simulations---}To numerically demonstrate the BKT RG flow for the Bose-Hubbard model, we perform high-precision numerical DMRG simulations for periodic chains of sizes up to $L=128$ sites using the \texttt{ITensors.jl} software library \cite{ITensor}. For OBC, we perform simulations up to $L=512$. 
We restrict the local site occupancy of bosons by $n_{\rm max}=6$---larger than those typically used, which may not be sufficient when considering fluctuations \cite{Barghathi2022}. 
Due to the non-periodic nature of the matrix product states used for the DMRG algorithm, the description of periodic systems requires large bond dimensions, which makes simulating large periodic systems challenging. Therefore, we carefully ensure the convergence of the DMRG algorithm by the following criteria: (i) no change in energy larger than $10^{-12}$ for consecutive sweeps, (ii) no change in the bond dimension for at least four sweeps using a truncation cutoff of $10^{-12}$, and (iii) a final energy variance satisfying $\ex{(H-E)^2}/E^2 < 10^{-11}$.

We perform DMRG calculations for many different values of the interaction $U/J$ in both the superfluid and Mott insulating phase for different system sizes $L$. Next, we compute the particle number fluctuations $\mathcal{F}$ as a function of the subsystem size $\ell$ in the ground state. We then extract the Luttinger parameter as described below Eq.~\eqref{Eq:Fpbc} (see Fig.~\ref{Fig:F_of_K_fit}) for each value of $U/J$ and $L$. We linearly fit to the region $\ell \in [L/2 - \lfloor L/16\rfloor, L/2]$ and obtain an error bar by varying the number of included points up to $\ell \in [L/4, L/2]$.   As a consistency check, a similar procedure is performed using ground state lattice worm quantum Monte Carlo with details provided in the supplement \cite{Supplement}

For OBC,  the bipartite fluctuations in Eq.~\eqref{Eq:Fpbc} computed for the Luttinger model, depends on the location of the bipartition \cite{Supplement}. Near the center of the interval, the PBC and OBC expressions agree with each other. However, in the OBC case, the Luttinger parameter $K$ acquires a strong spatial dependence due to the formation of image vortices at the boundaries \cite{Artemov:20052d}. This is in striking contrast to the assumption of a spatially constant $K$ used in the derivation of the LL prediction for bipartite fluctuations, which is valid only for a very small center interval in the presence of OBC. 
Extracting physically consistent values of $K$ requires extensive fine-tuning of this region, which may be responsible for the scatter of previously reported critical points discussed in the introduction. 
For this reason, PBC offer advantages for studying the phase transition, despite their increased computational cost in DMRG. 

Finally, approximate infinite matrix product state calculations can be performed directly in the
thermodynamic limit using the \texttt{ITensorInfiniteMPS.jl} software
implementing the VUMPS algorithm \cite{Zauner.2018VUMPS}.  We find excellent
agreement with the infinite system extrapolation of $K$ (see open circles in
Fig.~\ref{Fig:RGflowPBCdata}) deep in the superfluid phase where $V=0$.
However, close to the phase transition there is a discrepancy between the
infinite matrix product state approach and that extrapolated from finite $L$, which is plausible as the growth in entanglement \cite{Barghathi2022} cannot be captured at finite bond dimension \cite{Yang:2020td}. 

\begin{figure}[t!]
    \centering
    \includegraphics[width=\linewidth]{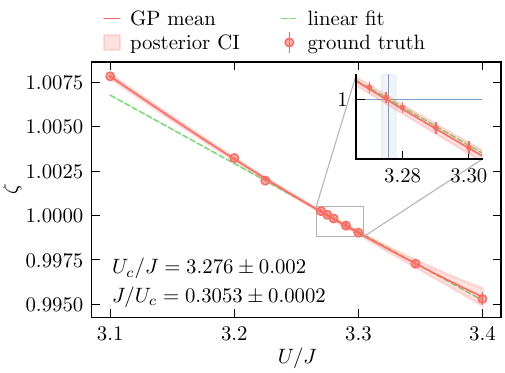}
    \caption{Dependence of the flow parameter $\zeta$ on the interaction strength $U/J$ (ground truth, red circles).  A Gaussian process (GP) is used to perform a non-parametric fit to the data, with the predicted posterior mean shown as a solid red line, accompanied by a red shaded region indicating the posterior confidence interval (CI).  The critical point occurs when $\zeta=1$ yielding $U_c/J=3.276\pm0.002$ indicated by the vertical blue line, and a shaded region of uncertainty in the inset.  Near the transition, a linear fit to $U/J\in\{3.275, 3.280, 3.290, 3.300\}$ (green dashed line) yields the same critical point and uncertainty, but away it deviates outside of error bars.\label{Fig:ZetaFit}}
\end{figure}

\emph{Demonstration of BKT flow---} For each $U/J$, the extracted values of $K$ for PBC at two consecutive system sizes are used in conjunction with Eq.~\eqref{Eq:RGEquation} to determine a value of $\zeta(U/J)$. Then, Eq.~\eqref{Eq:RGZeta} gives the associated value of $V$ corresponding to the $K$ of the smaller system size. Using the two largest values of $L$, this process gives the flow lines shown in Fig.~\ref{Fig:RGflowPBCdata} as well as their $V$-values.  We obtain error bars by additionally computing $\zeta$ using values of the Luttinger parameters varied within their error bars and taking the maximum deviation.   Thus, the observed agreement between flow lines and data points for smaller system sizes represents a stringent consistency check of BKT flow. Moreover, near the transition, we find excellent agreement with the predicted analytic scaling forms in Eqs.~\eqref{Eq:ScaleSF}-\eqref{Eq:ScaleINS} as seen in Fig.~\ref{Fig:K_of_L_FSS}.

Reconstruction of the RG flow allows us to precisely locate the phase transition at $U_c$ for which the flow with $\zeta(U_c/J)=1$ follows the separatrix in Fig.~\ref{Fig:RGflowPBCdata}.  
To extract the critical point, we plot $\zeta(U/J)$ in Fig.~\ref{Fig:ZetaFit} and employ a Gaussian process (GP) model and identify where the posterior mean predicts $\zeta(U_c/J)=1$, yielding $U_c/J=3.276(2)$ or $J/U_c =
0.3053(2)$. Uncertainty quantification is enabled via the posterior confidence interval of the GP \cite{Supplement}. Near the transition, we empirically observe $\zeta \propto U/J$, with large deviations outside it. 

{\em Conclusions}---Finite-size scaling is derived from the renormalization group (RG), making it essential to obtain numerical data consistent with a given  RG description. For the Berezinskii-Kosterlitz-Thouless transition, the RG framework for open boundary conditions is qualitatively different and significantly more complex than that for periodic boundary conditions. To ensure a robust data analysis, we have therefore obtained numerical data for  periodic boundary conditions. We observed finite-size scaling behavior at the quantum phase transition in the one-dimensional Bose-Hubbard model consistent with  the standard BKT RG flow, and obtained a high-precision estimate for the critical interaction strength at unit filling. Knowing its precise value may be 
important to quantify the effects of gapless modes and particle number fluctuations on entanglement \cite{Metlistski:2011ee,Kulchytskyy:2015gm,Barghathi2022}.  
We expect this approach to be useful in pinpointing critical points in other microscopic models where vortex proliferation is the dominant driver of the quantum phase transition. 

\emph{Code Availability} -- All software and data used in this study are available online \cite{paperrepo,datarepo}.

\acknowledgements 
This work was partially supported by the National Science Foundation Materials Research Science and Engineering Center program through the UT Knoxville Center for Advanced Materials and Manufacturing (DMR-2309083). 
M.T.~and B.R.~acknowledge support by DFG grant number RO 2247/16-1, and 
H.R.~acknowledges AITennessee for financial support. Computations were performed using resources provided by the Leipzig University Computing Center and University of Tennessee Infrastructure for Scientific Applications and Advanced Computing (ISAAC). 

%

    \ifarXiv
    

\section*{Supplementary material}

# Supplemental Material for: "Berezinskii-Kosterlitz-Thouless Renormalization Group Flow at a Quantum Phase Transition"

Matthias Thamm,[1] Harini Radhakrishnan,[2,3] Hatem Barghathi,[2,3] Chris Herdman,[4] Arpan Biswas,[5] Bernd Rosenow,[1] and Adrian Del Maestro[2,3,6]

[1] *Institut für Theoretische Physik, Universität Leipzig, 04103 Leipzig, Germany*
[2] *Department of Physics and Astronomy, University of Tennessee, Knoxville, TN 37996, USA*
[3] *Institute for Advanced Materials and Manufacturing, University of Tennessee, Knoxville, Tennessee 37996, USA*
[4] *Department of Physics, Middlebury College, Middlebury, VT 05753, USA*
[5] *University of Tennessee-Oak Ridge Innovation Institute, Knoxville Tennessee 37996, USA*
[6] *Min H. Kao Department of Electrical Engineering and Computer Science, University of Tennessee, Knoxville, TN 37996, USA*
(Dated: February 24, 2025)

## I. FLUCTUATIONS FOR PERIODIC BOUNDARY CONDITIONS

For a system of length $L$ with periodic boundary conditions (PBC), the boson field operators are given by [S1–S4]

$$\phi_{\mathrm{pbc}}(x) = \phi_0 + \pi x \frac{N}{L} - \frac{\mathrm{i}\pi}{L} \sum_{q \neq 0} \sqrt{\frac{L|q|}{2\pi}} \frac{1}{q} e^{-\alpha|q|/2 - \mathrm{i}qx} \left( b_q^\dagger + b_{-q} \right) \tag{S1}$$

$$\equiv \phi_0 + \pi x \frac{N}{L} + \tilde{\phi}_{\mathrm{pbc}} \tag{S2}$$

$$\theta_{\mathrm{pbc}}(x) = \theta_0 + \pi x \frac{J}{L} + \frac{\mathrm{i}\pi}{L} \sum_{q \neq 0} \sqrt{\frac{L|q|}{2\pi}} \frac{1}{|q|} e^{-\alpha|q|/2 - \mathrm{i}qx} \left( b_q^\dagger - b_{-q} \right) \quad, \tag{S3}$$

where $b_q$ ($b_q^\dagger$) are bosonic annihilation (creation) operators, $\phi_0$, $\theta_0$ are zero modes and $\alpha$ is a short-range interaction cutoff. We consider the density to leading order in the fields [S5] $\rho_{\mathrm{pbc}}(x) \approx \partial_x \phi_{\mathrm{pbc}}(x)/\pi \approx \rho_0 + \partial_x \tilde{\phi}_{\mathrm{pbc}}/\pi$ and for the purpose of computing the fluctuations, we neglect the $\cos(2\phi_{\mathrm{pbc}})$ term in the Luttinger Hamiltonian, main text Eq. (1). Defining $\tilde{\phi}_{\mathrm{pbc}}$ allows us to explicitly separate the density background $\rho_0$ from the fluctuations.

We can express the particle number operator by explicitly computing the integral of the density operator over the subregion $A \in [0, \ell]$ as

$$\hat{N}_A = \ell \rho_0 + \frac{1}{\pi} [\tilde{\phi}_{\mathrm{pbc}}(l) - \tilde{\phi}_{\mathrm{pbc}}(0)] \quad. \tag{S4}$$

In terms of the particle number operator, the fluctuations are given by

$$\mathcal{F}_{\mathrm{pbc}}(\ell; K, \alpha) = \frac{1}{\pi^2} \langle [\tilde{\phi}_{\mathrm{pbc}}(\ell) - \tilde{\phi}_{\mathrm{pbc}}(0)]^2 \rangle \quad. \tag{S5}$$

Under our assumptions, diagonalizing the Hamiltonian is equivalent to a rescaling of the boson field operators $\tilde{\phi}_{\mathrm{pbc}}(x) \to \sqrt{K}\tilde{\phi}_{\mathrm{pbc}}(x)$, $\tilde{\theta}_{\mathrm{pbc}}(x) \to \sqrt{K}^{-1}\tilde{\theta}_{\mathrm{pbc}}(x)$ [S5]. This allows us to explicitly compute the expectation value, yielding the expression for the fluctuations stated in the main text $\mathcal{F}_{\mathrm{pbc}} = \frac{K}{2\pi^2} \ln \left[ 1 + \frac{\sin^2(\pi\ell/L)}{\sinh^2(\pi\alpha/L)} \right]$.

## II. FLUCTUATIONS FOR OPEN BOUNDARY CONDITIONS

In the case of open boundary conditions (OBC), the displacement field can be written as

$$\phi_{\mathrm{obc}}(x) = \phi_B + \frac{\pi x}{L} N + \mathrm{i} \sum_{q>0} \sqrt{\frac{\pi}{qL}} e^{-\alpha q/2} \sin(qx) [b_q - b_q^\dagger] \tag{S6}$$

$$\equiv \phi_B + \frac{\pi x}{L} N + \tilde{\phi}_{\mathrm{obc}}(x) \quad, \tag{S7}$$

where $\phi_B \neq n\pi$ is a constant, such that the density vanishes at the ends of the chain [S3].

Here, the density operator including oscillating terms, is given by [S5]

$$\rho_{\text{obc}}(x) = \partial_x \phi_{\text{obc}}(x) \sum_{n=-\infty}^{\infty} \delta(\phi_{\text{obc}}(x) - n\pi) = \frac{1}{\pi} \partial_x \phi_{\text{obc}}(x) \sum_{p=-\infty}^{\infty} e^{-i2p\phi_{\text{obc}}(x)} \ . \tag{S8}$$

## A. Leading order fluctuations

To leading order, $m = n = 0$, we can compute the particle number fluctuations in an interval $A = [x, x + \ell]$ of length $\ell$ in analogy to the periodic case. We find

$$\mathcal{F}_{\text{obc}}(\ell; x) = \frac{1}{\pi^2} \sum_{q>0} \frac{\pi K}{L} \frac{1}{q} e^{-\alpha q} \left[\sin(q(x+\ell)) - \sin(qx)\right]^2 \tag{S9}$$

$$= \frac{K}{\pi^2} \ln \left| \frac{\sin\left[\frac{\pi}{L}\left(\frac{\ell}{2} - \frac{i\alpha}{2}\right)\right] \sqrt{\sin\left[\frac{\pi}{L}\left(x + \ell - \frac{i\alpha}{2}\right)\right] \sin\left[\frac{\pi}{L}\left(x - \frac{i\alpha}{2}\right)\right]}}{\sin\left[\frac{\pi}{L}\left(x + \frac{\ell}{2} - \frac{i\alpha}{2}\right)\right] \sinh\left[\frac{\pi\alpha}{2L}\right]} \right| \ . \tag{S10}$$

If we put the interval at the left boundary $x = 0$, the thermodynamic limit, $L \to \infty$, expression of $\mathcal{F}_{\text{obc}}$ does not agree with that for the periodic case, and we would not expect the RG flow to be the same. However, if we center the interval $A$ in the wire, choosing $x = (L - \ell)/2$, we fully recover the asymptotic form of the fluctuation formula from the periodic case

$$\mathcal{F}_{\text{obc}}\left(\ell; x = \frac{L-\ell}{2}\right) = \frac{K}{2\pi^2} \ln\left(1 + \frac{\sin^2(\pi\ell/L)}{\sinh^2(\pi\alpha/L)}\right) \approx \frac{K}{\pi^2} \ln \sin \frac{\pi\ell}{L} + A \ . \tag{S11}$$

In contrast, if one considers the interval $[0, \ell]$ starting at the left edge, the fluctuations become

$$\mathcal{F}_{\text{obc}}(\ell; x = 0) = \frac{K}{4\pi^2} \ln\left(1 + \frac{\sin^2(\pi\ell/L)}{\sinh^2(\pi\alpha/2L)}\right) \ . \tag{S12}$$

In the thermodynamic limit, this result differs from the periodic case, as $x = 0$ for $L \to \infty$ corresponds to a semi-infinite chain.

## B. Including higher order terms

To leading order $m = n = 0$, the result does not capture boundary effects such as Friedel oscillations that are present in an open chain. All terms $n, m$ can be computed analytically using

$$\rho(x) = \frac{-i}{2\pi} \sum_{n=-\infty}^{\infty} \partial_x \left[\frac{1}{n} e^{2ni(\pi\rho_0 x + \phi(x))}\right] \ , \tag{S13}$$

and interchanging the derivative with the sum before integrating the density. By introducing Green functions $G_0(x) = \langle \tilde{\phi}_{\text{obc}}^2(x)\rangle \equiv \ln F_0(x)$ and $G_1(x,y) \equiv \ln F_1(x,y) = \langle \tilde{\phi}_{\text{obc}}(x)\tilde{\phi}_{\text{obc}}(y)\rangle$, we can express the fluctuations as

$$\mathcal{F}(\ell; x) = \frac{-1}{4\pi^2} \sum_{nm} \frac{1}{nm} e^{2i(n+m)(\pi\rho_0 x + \phi_B)} \left[e^{2(m+n)i\pi\rho_0\ell}\left(F_0^{-2(n+m)^2}(x+\ell) - F_0^{-2(m^2+n^2)}(x+\ell)\right)\right.$$
$$\left. + F_0^{-2(m+n)^2}(x) - F_0^{-2(m^2+n^2)}(x) - 2e^{2mi\pi\rho_0\ell}F_0^{-2m^2}(x+\ell)F_0^{-2n^2}(x)\left(F_1^{-4nm}(x+\ell,x) - 1\right)\right] \ . \tag{S14}$$

One can explicitly compute the Green functions, yielding

$$F_0(x) = \left(1 + \frac{\sin^2\left(\frac{\pi x}{L}\right)}{\sinh^2\left(\frac{\pi\alpha}{2L}\right)}\right)^{\frac{K}{4}} \tag{S15}$$

$$F_1(x,y) = \left(\frac{\cos\left(\frac{\pi}{L}(x+y)\right) - \cosh\left(\frac{\pi\alpha}{L}\right)}{\cos\left(\frac{\pi}{L}(x-y)\right) - \cosh\left(\frac{\pi\alpha}{L}\right)}\right)^{\frac{K}{4}} \ . \tag{S16}$$

However, we find that including the higher order terms has effectively no effect on the fluctuations, as the $n = m = 0$ term dominates the expression. We therefore do not include them in the comparison below.

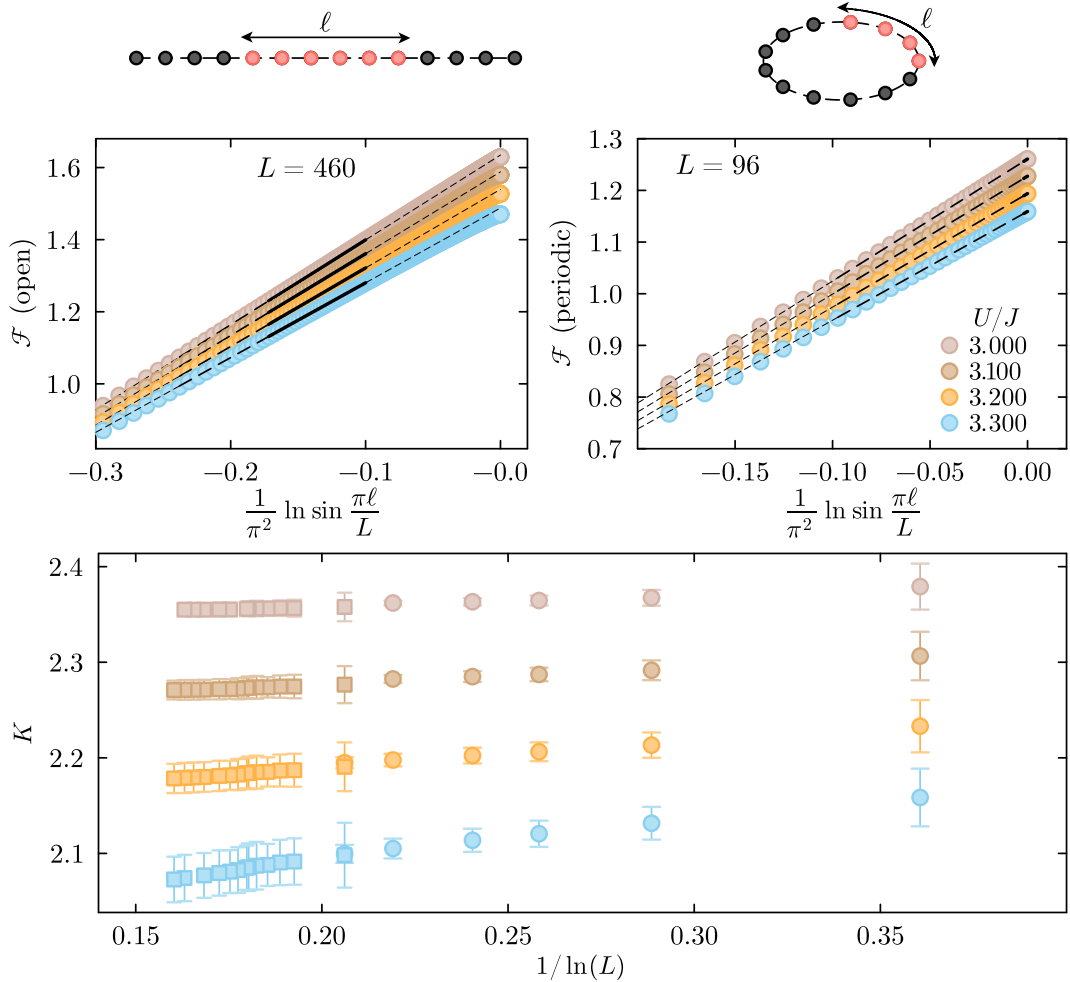

FIG. S1. Comparison of particle number fluctuations for Bose-Hubbard chains with periodic (top left panel) and open boundary conditions (top right panel). We show the fluctuations $\mathcal{F}$ for partition sizes $\ell$ up to $L/2$ for several values of the interaction strength $U/J$ with linear fits (solid black) over regions described in the main text and extended (dashed black). The bottom panel shows the flow of $K$ depending on the system size $L$, with a good agreement between PBC (circles) and OBC results (squares), owing to the fine-tuned OBC fitting method illustrated in Fig. S2.

## III. CHALLENGES WITH OPEN BOUNDARIES

In Fig. S1, we show a comparison of extracting the Luttinger parameter $K$ from particle number fluctuations $\mathcal{F}$ for the cases of periodic vs open boundary conditions (PBC vs. OBC), and demonstrate the system size dependence of $K$. For PBC, we perform DMRG calculations up to $L = 128$ while for OBC, we can study larger system sizes, up to $L = 512$ sites. Error bars in the extracted value of $K$ are determined by systematically varying the fit region. For PBC, the linear fits are mostly insensitive to this variation leading to small error bars. For OBC, on the other hand, we observe large transients and boundary effects that are relevant even for $L = 460$ (top right panel), where $\mathcal{F}$ can be seen to bend away from the predicted linear behavior for $\ell \simeq L/2$. This effect can be seen in more detail in Fig. S2 top panels where the maximum size of the fitting region is adjusted from $L/6$ to $L/10$ leading to large deviations in the extracted value of $K$, where different fits may produce $K$-values lying outside a standard deviation of each other.

To avoid both the UV effects for small $\ell$ and the significant boundary effects for $\ell \sim L/2$, we optimize a fit interval and find $L/8.2$ for OBC which minimizes both of these effects and maximizes consistency with the PBC case as seen in the lower panels of Fig. S2. Using this diminutive fit window allows us to achieve agreement of the OBC scaling with PBC, as well as reproduce the BKT flow (center panel) corresponding to Eqs. (4)–(6) in the main text. However, the resulting large error bars do not allow us to accurately pinpoint the critical $U_c/J$ from the OBC data for the system sizes we can study with DMRG.

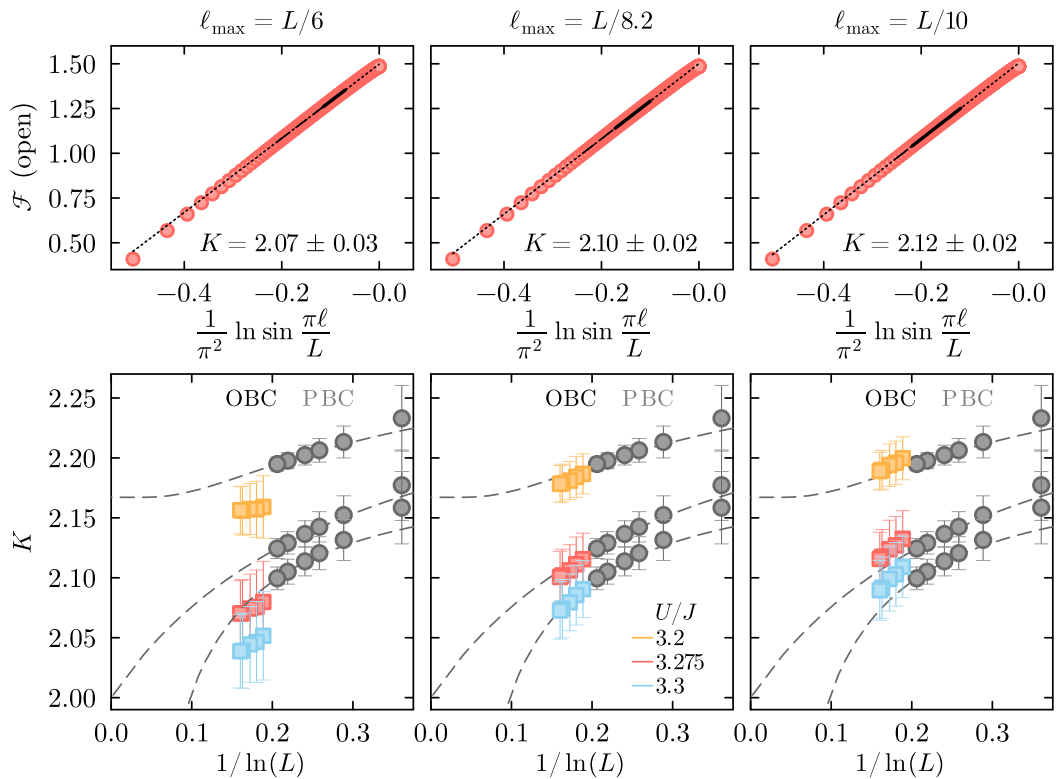

FIG. S2. Extracting the Luttinger parameter $K$ from the particle number fluctuations for different fit intervals to the DMRG data for open boundary conditions (OBC). We fit Eq. (S11) to the OBC DMRG data at $L = 460$ (upper panels) for intervals $\ell \in [\ell_{max} - L/16, \ell_{max}]$ with $\ell_{max} = L/6$ (left), $\ell_{max} = L/8.2$ (center), $\ell_{max} = L/10$ (right). The lower panels depict the corresponding finite size scaling of the Luttinger parameter. Analogous to main text Fig. 2, we show RG flows from the superfluid phase, separatrix, and insulating phase. The gray points are the results for periodic boundaries as presented in the main text. Squares show the open boundary values of $K$ obtained with the corresponding fit intervals. In contrast to the periodic case, we find strong dependence of the Luttinger parameters from the fitting interval for open boundaries. In the main text, we use the fine-tuned fit interval depicted in the center panels.

Previous OBC work [S6] studied fluctuations in the interval $[0, \ell]$, thus including a boundary. In this case, the vortex fugacity depends on the distance from the boundary, and via the RG flow, the Luttinger parameter $K$ acquires a position dependence as well [S5, S7]. As a consequence, the relation Eq. (S10) (obtained for a spatially constant $K$) is no longer valid, and extracting $K$ from particle number fluctuations is subtle. In Fig. S3 (left panel) we reproduce the fitting method reported in Ref. S6 along with our inferred finite size scaling (right panel) for the fluctuations in $[0, \ell]$. To achieve this, we let $x = 0$ in Eq. (S10), resulting in Eq. (S12)). The example fit ($L = 128, U/J = 3.3333$) shows excellent agreement with the one reported in the supplement of Ref. S6 and a $1/\ln L$ finite size scaling analysis for the three red points $L \in [128, 256, 512]$ used in Ref. S6 confirms that their critical interaction strength of $U/J = 3.3456$ indeed extrapolates as $K \to 2$. However, when analyzing the system sizes in between (blue points, generated by us), strong deviations are observed, highlighting that strong OBC effects make the use of Eq. (9) in the main text problematic.

## IV. GROUND STATE QUANTUM MONTE CARLO

As a second consistency check of our PBC DMRG, we performed quantum Monte-Carlo (QMC) simulations using the lattice worm path integral Monte-Carlo algorithm `pigsfli` [S8]. These simulations are challenging and resource intensive due to the requirement of a high precision estimation of the fluctuations in the region $\ell \sim L/2$, where the Luttinger theory is most accurate but the QMC has the largest statistical fluctuations. This is exacerbated by polynomial imaginary time $\beta$-scaling in presence of a zero mode. We include the QMC data in the reconstructed flow diagram (crosses in Fig. 1 of the main text), which are in excellent agreement with the flow obtained from PBC DMRG.

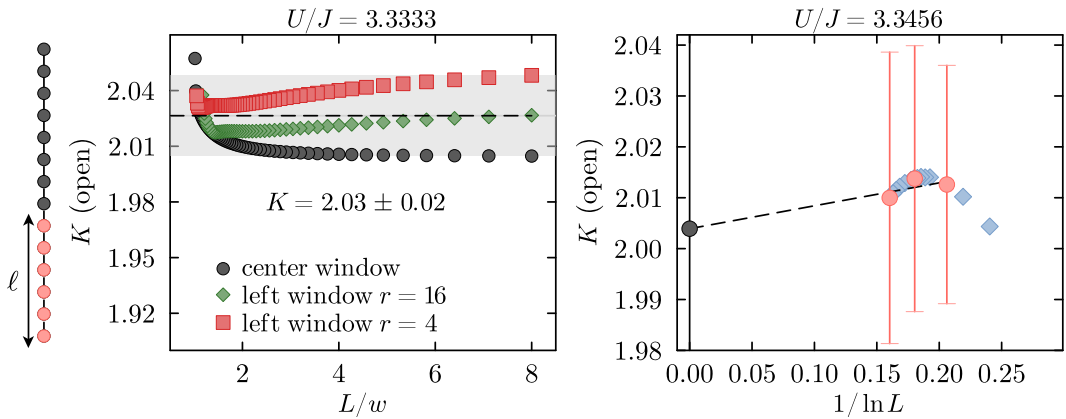

FIG. S3. Reproduction of the fitting method (left panel) and our inferred finite size scaling (right panel) of Ref. S6 for open boundary condition fluctuations in region $A = [0, \ell]$. The left panel depicts the dependence of the extracted Luttinger parameter when fitting to intervals $\ell \in [4, w+4]$ (red squares), $[16, 16+w]$ (green diamonds), and $[L/2 - w/2, L/2 + w/2]$ (black circles) on the size $w$ of the fitting window for $L = 128, U/J = 3.3333$. The Luttinger parameter is then estimated as the mean between center and left $(r = 4)$ interval at $L/w = 8$. Our left panel is in excellent agreement with the inset in the right panel of Fig. 2 in the supplement of Ref. S6. The right panel shows finite size scaling of the Luttinger parameter for the points of Ref. S6 (red) and more system sizes in between (blue) at the critial point $U/J = 3.3456$ reported in Ref. S6. While scaling as $1/\ln(L)$ to $L \to \infty$ only from the red points yields a $K(L \to \infty) \approx 2$ within error bars (black circle), the finite size scaling including many system sizes is inconsistent with extrapolation from the red points only. This inconsistency of extrapolation is reflected by the large error bars we obtain for the red points. We attribute these observations to the strong boundary effects fully present in the $A = [0, \ell]$ region for open boundary conditions.

## V.   ESTIMATION OF THE CRITICAL POINT VIA BAYESIAN OPTIMIZATION

The simultaneous solution of the RG equations (2)–(3) in the main text allows us to numerically evaluate $\zeta(U/J)$ in the thermodynamic limit for the Bose-Hubbard model through the dependence $K(U/J)$. The critical coupling $U_c/J$ can then be determined by identifying when $\zeta(U_c/J) = 1$. However, as described throughout this work, obtaining a reliable and accurate value of $K(U/J)$ is costly, requiring the use of PBC simulations (either DMRG or QMC) with considerable computational effort. A precise grid search is thus not feasible. We also have no access to analytical derivatives of this function, limiting the ability to use gradient based approaches.

While we empirically observe that near the transition $\zeta \propto U/J$, deviations outside the critical region are large, and an assumption of linearity could introduce bias and/or systematic error to the extraction of $U_c$. To mitigate the need for any assumptions on the functional form of $\zeta(U/J)$ and reduce the computational cost of identifying $U_c/J$, we employ a surrogate non-parametric Gaussian process (GP) model [S9] for $\zeta$:

$$\zeta(\boldsymbol{x}) \sim \mathrm{GP}(\mathbb{E}[\zeta(\boldsymbol{x})], \mathrm{cov}(x, x')) \tag{S17}$$

with zero mean function $\mathbb{E}[\zeta(\mathbf{x})] = 0$ and correlations between the ground truth data, $x_n \equiv (U/J)_n$ along with their fixed uncertainties $\Delta(U/J)_n$, are expressed through a covariance matrix computed using a Matern kernel function with smoothness parameter $\nu = 5/2$ (chosen via manual tuning):

$$\mathrm{cov}(x, x') = \theta_0 \left(1 + \sqrt{5} \frac{|x - x'|}{\theta_1} + \frac{5}{3} \frac{(x - x')^2}{\theta_1^2}\right) \exp\left(-\sqrt{5} \frac{|x - x'|}{\theta_1}\right) . \tag{S18}$$

The scale $(\theta_0)$ and length $(\theta_1)$ hyperparameters were optimized via a gradient based method, and the GP model was fit using the `GPyTorch` library [S10] using $x_n \in \{3.1, 3.2, 3.225, 3.27, 3.28, 3.3, 3.346, 3.4\}$. The results, including multiple draws from the posterior distribution and the resulting GP mean, are shown in the left panel of Fig S4. We identify the critical point as the value of $U/J$ where the predictive mean of the GP $\mu = 1$ and its associated uncertainty can be quantified through the confidence interval corresponding to a single standard deviation $\sigma$ of the surrogate model.

To understand the robustness of this prediction, we transform our data via $y = (1 - \zeta)^2$ for the same inputs as shown in the middle panel of Fig. S4. This allows us to recast finding the critical point as a convex optimization problem that can be accelerated via sample efficient Bayesian optimization (BO) methods [S11–S13]. In particular, we employ the combined power of the `GPyTorch` and `BoTorch` [S14] software libraries using using the same zero mean and covariance function as in Eq. (S18). Implementation details are included in an associated repository [S15]. To

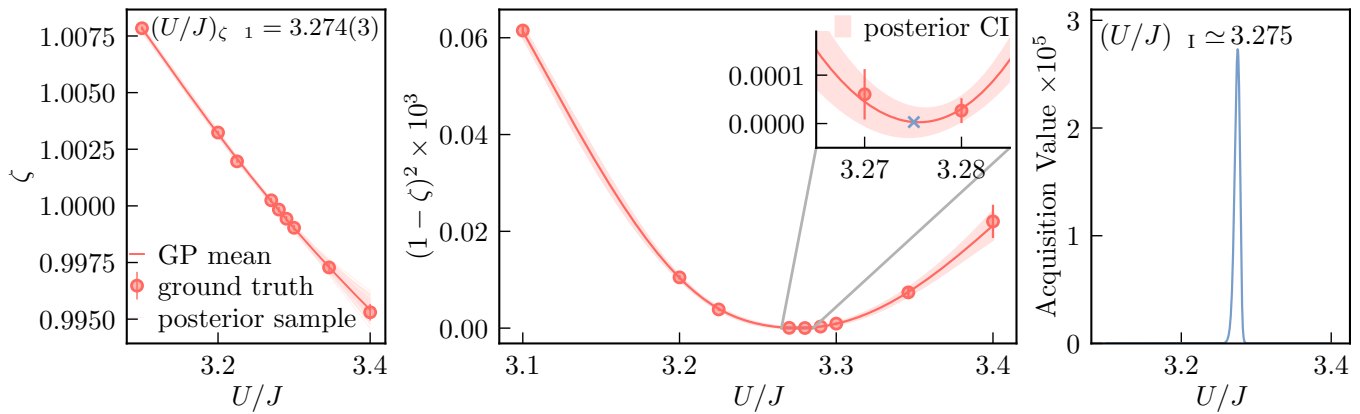

FIG. S4. Gaussian process (GP) surrogate and Bayesian optimization (BO) approach to determine the critical point. Left: Data points represent ground truth values for $\zeta$ while the faint pink lines are 100 draws from the posterior distribution for the GP model in Eq. (S18) along its associated mean. The critical point can be identified by determining the value of $U/J$ which corresponds to $\zeta = 1$ as shown in the top of the panel along with its associated uncertainty in the last digit shown in parenthesis. Middle: a fit GP model for the transformed data $(1 - \zeta)^2$ allowing us to recast finding the critical point as an optimization problem. The blow-up in the inset shows the posterior distribution confidence interval (CI) as well as a × at the parametric value of $U/J$ where the expected improvement (EI) function is maximized (right panel).

choose the next sample guided by BO, we used the Expected Improvement (EI) acquisition function

$$\text{EI}(x) = (\mu(x) - y_{\text{best}} - \epsilon)\Phi\left(\frac{\mu(x) - y_{\text{best}} - \epsilon}{\sigma(x)}\right) + \sigma(x)\phi\left(\frac{\mu(x) - y_{\text{best}} - \epsilon}{\sigma(x)}\right) \tag{S19}$$

where $\mu_y(x)$ and $\sigma_y(x)$ are the GP predictive mean and standard deviation, $y_{\text{best}}$ is the optimal value of the target in the training set, $\Phi(\cdot)$ is the cumulative normal distribution function, and $\phi(\cdot)$ is the normal probability distribution function. A slack parameter $\epsilon = 0.01$ was added for numerical stability and balancing between exploration and exploitation [S16]. A single BO iteration identified the optimal next sampling point as $U/J = 3.275$ via a localized maximum of the acquisition function shown in the right panel of Fig. S4. The small peak value $\text{EI} \approx 3 \times 10^{-5}$ signifies near convergence (rare likelihood that any new better/improved solution can found than the last suggested). We then performed one additional set of DMRG calculations at the identified $U/J = 3.275$ allowing us to pinpoint the critical point with a small and well quantified error $\sim \text{O}(10^{-3})$ as shown in Figure 4 of the main text.

---


    \fi

\end{document}